\magnification=1200
\parindent=0truecm \parskip=0truecm
\baselineskip=0.82truecm \parindent=1.2truecm
\parskip=0.25truecm
\hsize=15.7truecm
\vsize=22.5truecm

\centerline{NOVEL TYPE OF PHASE TRANSITION IN A SYSTEM }
\centerline{OF SELF-DRIVEN PARTICLES }
\vskip0.9truecm

\centerline { Tam\'as Vicsek$,^{a,b}$  Andr\'as Czir\'ok,$^{a}$
Eshel Ben-Jacob,$^c$  Inon Cohen,$^c$} \smallskip
\centerline{ and Ofer Shochet,$^c$ }

\bigskip
\parskip=0.0truecm

\item{$^a$} Department of Atomic Physics, E\"otv\"os University,
Budapest, Puskin u 5-7, 1088 Hungary

\item{$^b$} Institute for Technical Physics, Budapest, P.O.B. 76, 1325
Hungary

\item{$^c$} School of Physics, Tel-Aviv University, 69978 Tel-Aviv,
Israel

\parskip=0.3truecm

\vskip1truecm

\centerline{\bf Abstract}

A simple model with a novel type of dynamics is introduced in order to
investigate the emergence of self-ordered motion in systems of particles
with biologically motivated interaction.  In our model particles are
driven with a constant absolute velocity and at each time step assume
the average direction of motion of the particles in their neighborhood
with some random perturbation ($\eta$) added.  We present numerical
evidence that this model
results in a kinetic phase transition from no transport (zero average
velocity, $\vert {\bf v}_a \vert =0$) to finite net transport through
spontaneous symmetry breaking of the rotational symmetry.  The
transition is continuous since $\vert {\bf v}_a \vert$ is found to scale
as $(\eta_c-\eta)^\beta$ with $\beta\simeq 0.45$.

\bigskip

\centerline{--------------------}

\bigskip

One of the most interesting aspects of many particle systems is that
they exhibit a complex cooperative behavior during phase transition [1].
This remarkable feature of equilibrium systems has been studied in great
detail for the last couple of decades leading to a deeper understanding
of
processes which may take place in an assembly of interacting
particles. Concepts like scaling, universality and renormalization
have resulted in a systematic picture of a wide range of systems in
physics [1,2].

Recently there has been an increasing interest in the rich behaviour of
systems which are far from equilibrium.  Processes such as aggregation,
viscous flows or biological pattern formation have been shown to involve
scaling of the related geometrical and dynamic quantities characterizing
these phenomena [3,4].  As a further similarity with equilibrium
systems, the
existence of phase transition type behavior has also been demonstrated
in several investigations of growth  processes [5-8].  These analogies
with the basic features
of equilibrium systems have represented a particularly important
contribution to the understanding of the complex behavior of
non-equilibrium processes.

In this work we introduce a model with a novel type of dynamics in order
to investigate clustering, transport and phase transition in
non-equilibrium systems where the {\it velocity} of the particles is
determined by a simple rule and random
fluctuations. {\it The only rule of the model is: at each time step a
given particle driven with a constant absolute velocity assumes the
average direction of motion of the particles
in its neighborhood of radius $r$ with some random perturbation added.}
We show using
simulations that in spite of its simplicity, this model results in a
rich, realistic dynamics, including a kinetic phase transition
from no transport to finite net transport through  spontaneous
symmetry breaking of the rotational symmetry.

In this sense our model is a {\it transport related,
non-equilibrium analog of the ferromagnetic
type of models}, with the important difference that it is inherently
dynamic: the {\it elementary event is the motion of a particle} between
two time steps.  Thus, the analogy can be formulated as follows: the
rule corresponding to the ferromagnetic interaction tending to align the
spins in the same direction in the case of equilibrium models is
replaced
by the rule of {\it aligning the direction of motion} of particles in
our model of cooperative motion. The level of random perturbations we
apply are in  analogy with the temperature.

Beyond the above aspects (analogies and simplicity) the proposed model
is interesting because of possible applications in a wide range of
biological systems involving clustering and migration.  Biological
subjects have the tendency to move as other subjects do in their
neighborhood [9].  In addition to such trivial examples as schools of
fish, herds of quadrupeds or flocks of flying birds, our model can be
applied to the less known phenomena during bacterial colony growth [10].
There are bacteria (e.g., a strain of {\it Bacillus Subtilis}) which
exhibit cooperative motion in order to survive under unfavorable
conditions.  The present model with some modifications is already
capable to reproduce the main observed features of the motion
(collective
roatation and flocking) of bacteria [10].  Other biologically
motivated, recent theoretical investigations of clustering, aggregation
and orientational order in systems with diffusing directed objects have
concentrated on the possible spatial patterns arising from an
integro-differential equation approach and from cellular automata type
models [11].

Furthermore, we expect that our model can be used to interpret the
results of experiments  on clustering and convection
in a system of disks floating on an air table [12].  These experiments
represent a physically motivated possible application of the present
model, since they are being carried out in order to understand the flow
of granular materials under specific conditions.  We are aware that two
groups are working on developing models similar to ours in order to
interpret these air table experiments [13].

The actual simulations were carried out in a square shaped cell of
linear size $L$ with
periodic boundary conditions.  The particles were represented by points
moving continuously (off-lattice) on the plane. We used the interaction
radius $r$ as unit to measure distances ($r=1$) while the time unit,
$\Delta t=1$ was the time interval between two updating of the
directions/positions.
 In most of our
simulations we used the simplest initial conditions: i) at time $t=0$
$N$ particles were randomly distributed in the cell and ii) had the same
absolute velocity $v$ and iii)  randomly distributed directions
$\theta$.
  The velocities $\{{\bf
v}_i\}$
of the particles were determined simultaneously at each time step and
the position of the $i$th particle updated according to
$$
{\bf x}_i(t+1)={\bf x}_{i}(t)+{\bf
v}_i(t)\Delta t.  \eqno(1)
$$
  Here the velocity of a particle ${\bf v}_i(t+1)$ was constructed
to have an absolute value $v$ and a direction given by the angle
$\theta(t+1)$. This angle
 was obtained from the
expression
$$
\theta(t+1)= \langle \theta(t)\rangle_r+\Delta \theta, \eqno(2)
$$
 where
$\langle \theta(t)\rangle_r$ denotes the average
direction of the velocities of particles (including  particle $i$)
being within a circle of radius $r$ surrounding the given particle. The
average direction was given by the angle
${\rm
arctg}[\langle\sin(\theta(t))\rangle_r/\langle\cos(\theta(t))\rangle_r]$.
In (2) $\Delta \theta$ is a random number chosen with a unifom
probability from the interval $[-\eta/2, \eta/2]$. Thus,
the term $\Delta \theta$ represents noise which we shall use as a
temperature-like variable.
Correspondingly, there are three free parameters for a
given system size: $\eta$, $\rho$ and $v$, where $v$ is the distance a
particle makes between two updatings.

We have chosen this realization because of its simplicity, however,
there may be several more interesting alternatives of implementing the
main rules of the model.  In particular, the absolute value of the
velocities does not have to be fixed, one can introduce further kind of
interaction between a particles and
or consider lattice alternatives of the model.  In the rest of this
paper we shall concentrate on the above described simplest version and
investigate the nontrivial behavior of the transport properties as the
two basic parameters of the model, the noise $\eta$ and the density
$\rho=N/L^2$ are varied. We used $v=0.03$ in the simulations we report
on
because of the following reasons. In the limit  $v\to 0$ the particles
do not move and the model becomes an analog of the well known XY model.
For $v\to \infty$ the particles become completely mixed between two
updates and this limit corresponds to the so called mean field behavior
of a ferromagnet. We use  $v=0.03$ for which the particles always
interact with their actual neighbours and move fast enough to change the
configuration after a few updates of the directions. According to our
simulations in a wide range of the velocities ($0.003 < v < 0.3$) the
actual value of $v$ does not affect the results.

Fig. 1a-d demonstrates the velocity fields during runs with various
selections for the value of  the parameters $\rho$ and $\eta$. The
actual velocity of a  particle is indicated by a small arrow, while
their
trajectory for the last 20 time steps is shown by a short
continuous curve. (a) At $t=0$ the positions
and the direction of velocities are distributed randomly.  (b) For small
densities and noise the particles tend to form groups moving coherently
in random directions.  (c) At higher densities and noise the particles
move randomly with some correlation.  (d) perhaps the most interesting
case is when the density is large and the noise is small; in this case
the motion becomes ordered on a macroscopic scale and all of the
particles tend to {\it move in the same spontaneously selected
direction}.

This kinetic phase transition is due to the fact that the particles are
driven with a constant absolute velocity; thus, unlike standard
physical
systems in our case the {\it net momentum of the interacting particles
is
not conserved} during collision.
We have studied in detail the nature
of this transition by determining the absolute value of the average
normalized velocity
$$
v_a={1\over Nv}\left\vert \sum_{i=1}^{N} {\bf v}_i \right\vert \eqno(3)
$$
 of the entire
system of particles as the noise and the density were changed.  This
velocity is approximately zero if the direction of the motion of the
individual particles is distributed randomly, while for the coherently
moving phase (with ordered direction of velocities) $v_a \simeq 1$, so
that we can consider the average velocity as an {\it order parameter}.

First we gradually decreased the amount of noise $\eta$ in cells of
various sizes for a fixed density $\rho$ and observed a transition from
a disorderly moving phase to a phase with a coherent motion of the
particles (Fig. 2a).  The uncertainity of the data points is within the
range of the symbols except for runs carried out
with 4000 and 10000
particles  close to the transition.  For these $\eta$ values the
statistical errors estimated from 5 runs with different initial
conditions are in the range of 5\% (resulting in an overlap of the
results for a limited number of $\eta$ values) due to the slow
convergence and large fluctuations.  In Fig.~2b we show how $v_a$
changes if the noise is kept constant and the density is increased.

Quite remarkably, the behavior of the kinetic order parameter $v_a$ is
very similar to that of the order parameter of some equilibrium systems
close to their critical point.  The strongest indication of a transition
in our nonequilibrium model is the fact that as we go to larger system
sizes the region over which the data show scaling is increasing (see
Fig. 3a).  Only an extremely unusual crossover could change this {\it
tendency}.  A plausible physical picture behind our finding is the
following: since the particles are diffusing, there is mixing in the
system resulting in an effective (long range) interaction radius.

Thus, we can assume that in the thermodynamic limit our model exhibits a
kinetic phase transition analogous to the continuous phase transition in
equilibrium systems, i.e.,
$$
v_a\sim (\eta_c(\rho)-\eta)^\beta
\hskip1.2truecm {\rm and} \hskip1.2truecm v_a\sim
(\rho-\rho_c(\eta)),^\delta \eqno(4)
 $$
where $\beta$ and $\delta$ are
critical exponents and $\eta_c(\rho)$ and $\rho_c(\eta)$ are the
critical noise and density (for $L\to\infty$), respectively.  We can
determine $\beta$ and $\delta$ corresponding to the rate of vanishing of
the order parameter from plotting $\log v_a$ as a function of $\log
[\eta_c(L)-\eta)/\eta_c(L)]$ and $\log [\rho-\rho_c(L)/\rho_c(L)]$
for some fixed
values of $\rho$ and $\eta$, respectively (Fig.~3).  For finite sizes
$\eta_c(L)$ and $\rho_c(L)$ are $L$ dependent, thus, we used such values
of quantities for which the plots in Fig.~3 were the straightest
in the relevant region of noise or density values.  The slope of the
lines fitted to the data can be associated with the critical exponents
for which we obtained $\beta= 0.45\pm 0.07$ and $\delta=0.35\pm 0.06$.
The errors in determining $\beta$ and $\delta$ are due to the
uncertainties in the i) $v_a$ and the ii) $\eta_c(L)$ and $\rho_c(L)$
values.  Since the scaling plots in Fig.~3 depend sensitively on the
choice
of the critical noise and density and our method of determing their
value is indirect (from the straightness of the data sets) we give
rather conservative estimates for the errors of $\beta$ and $\delta$.

 We have carried out a finite size scaling
analysis of $\eta_c(L)$ and obtained
$\eta_c(\infty)=2.9\pm 0.05$
for $\rho=0.4$ (note that the  ``infinite
temperature'' limit of our model is $\eta_c=2\pi$).  As indicated,
$\eta_c$
depends on $\rho$, in fact, we expect a phase diagram (a line of
critical temperatures) analogous to that of disordered ferromagnets,
$\eta_c$ playing the role of temperature and $\rho$ playing the role of
the density of spins.  In this case $\beta$ and $\delta$ are expected to
have the same value. On the other hand,
strong crossover effects are likely
to effect their actual values in a finite size simulation. Although
our estimates for $\beta$ and $\delta$ are different, on the
basis of our simulations we cannot exclude the possibility
(allowed by our error bars)
that they
become equal in the thermodynamic limit.
However,
the determination of the phase diagram and a more precise calculation of
the exponents of our new model is outside of the scope of the present
work which concentrates on demonstrating the main features of a novel
nonequilibrium system.

The emergence of cooperative motion in our model has analogies with the
appearance of spatial order in equilibrium systems.  This fact, and the
simplicity of our model suggests that, with appropriate modifications,
the theoretical methods for describing critical phenomena may be
applicable to the present kind of far-from-equilibrium phase transition.
The kinetic phase transitions which have been observed in surface growth
models [5-8] are both in analogy and different from the situation
described here.  The similarity is in the scaling behaviour of an
inherently non-equilibrium order parameter, while the two kinds of
processes are distinct from the point of the driving force acting on the
particles.  Self-driven particles are uncommon in physics, but they are
typical in biological systems, including live organisms and the so
called ``molecular motors'' having attracted great interest recently
[14].  Transitions have been observed in traffic models [15] consisting
of particles (cars) which can also be interpreted as self-driven
particles.

There are interesting further  variations of the model
investigated in this work. It is expected that taking into account
a hard core term in the interaction or using  semi periodic or open
boundary conditions results in additional non-trivial effects. Our
preliminary results [10] indicate that a model with hard core repulsion
and  specific boundary conditions can be successfully used to interpret
recent observations of coherent motions in geometrically
complex bacterial colonies growing on soft agar surfaces [10,16-18].

{\bf Acknowledgement}: The present research in part was supported by the
Hungarian Research Foundation Grant No. T4439, by a grant from the
German-Israeli Foundation for Scientific Research and Development, and
by the Program for Alternative Thinking at Tel-Aviv University.

\vfill\eject
\noindent {\bf  References}
\baselineskip=0.82truecm
\parindent=0truecm   \parskip=0truecm

\def\pp{\parshape=3 0truecm 15.7truecm 1truecm 14.7truecm
1truecm 14.7truecm}

\pp [1] H. E. Stanley {\it Introduction to Phase Transitions and
Critical Phenomena}  (Oxford University Press, Oxford, 1971)

\pp [2] see, e.g., S.-K.  Ma {\it Statistical Mechanics} (World
Scientific, Singapore, 1985); {\it Modern Theory of Critical Phenomena}
(Benjamin, New York, 1976)

\pp [3] P. Meakin,  in {\it Phase Transitions and Critical
Phenomena} Vol 12. edited by C. Domb and J. Lebowitz ( Academic Press,
New York, 1987)

\pp [4] T. Vicsek, {\it Fractal Growth Phenomena.} (Singapore, World
Scientific, 1992)

\pp [5] R. Jullien, J. Kert\'esz, P. Meakin and D. Wolf, eds., {\it
Surface Disordering.} (New York, Nova Science, 1992)

\pp [6] J. Kert\'esz, and D. E. Wolf,  {\it Phys.  Rev. Lett.} {\bf
62}, (1989) 2571

\pp [7] N. Martys, M. Cieplak, and M. O. Robbins,   {\it Phys.
Rev. Lett.} {\bf 66}, (1991) 1058

\pp [8] Z. Csahok, K. Honda, E. Somfai, M. Vicsek and T. Vicsek, {\it
Physica} {\bf A200}, (1993) 136

\pp [9] D. P. O'Brien {\it J. Exp. Mar. Biol. Ecol.} {\bf 128}, (1989) 1

\pp [10] E. Ben-Jacob, A. Czir\'ok, I. Cohen, O. Shochet, A. Tenenbaum
and T. Vicsek to be published

\pp [11] L. Edelstein-Keshet and G. B. Ermentrout, {\it J. Math. Biol.}
{\bf 29}, (1990) 33

\pp [12] J. Lemaitre, A. Gervois, H. Peerhossaini, D. Bideau and J. P.
Troadec, {\it J. Phys.  D: Appl.  Phys.} {\bf 23} (1990) 1396

\pp [13] H. J. Herrmann, private communication

\pp [13] see, e.g., R. D. Astumian and M. Bier {\it Phys.  Rev. Lett.}
{\bf 72}, (1994) 1766

\pp [14] see, e.g., K. Nagel and H. J. Herrmann, {\it Physica A} {\bf
199}, (1993) 254  and references therein

\pp
[15] E. Ben-Jacob, H. Shmueli, O. Shochet and A. Tenenbaum  {\it
Physica} {\bf A187}, (1992) 378

\pp [16] E. Ben-Jacob, I. Cohen, O. Shochet, A. Tenenbaum,
A. Czir\'ok,and T. Vicsek, {\it Nature} {\bf 368}, 46 (1994)

\pp [17] T. Matsuyama, R. M. Harshey and M. Matsushita, {\it Fractals}
{\bf 1}, 302 (1993)

\vfill \eject
\parindent=0truecm \parskip=0.4truecm
 {\bf Figure captions}

Figure 1.  In this figure the velocities of the particles are displayed
for varying values of the density and the noise.  The actual velocity of
a particle is indicated by a small arrow, while their trajectory for the
last 20 time step is shown by a short continuous curve.  The number of
particles is $N=300$ in each case.  (a) $t=0$, $L=7$ $\eta=2.0$ (b) For
small densities and noise the particles tend to form groups moving
coherently in random directions, here $L=25$, $\eta=0.1$; (c) After some
time at higher densities and noise ($L=7$ $\eta=2.0$) the particles move
randomly with some correlation; (d) For higher density and {\it small
noise} ($L=5$ $\eta=0.1$) the motion becomes ordered. All of our
results shown in Figs.~1-3 were obtained from simulations in which $v$
was set to be equal to 0.03.

Figure 2.  (a) The absolute value of the average velocity ($v_a$) versus
the noise $\eta$ in cells of various sizes
for a fixed density $\rho$. The symbols correspond to $\sqcap$ - $N=40$,
$L=3.1$; $+$ - $N=100$,  $L=5$; $\times$ - $N=400$, $L=10$;
$\bigtriangleup$ - $N=4000$, $L=31.6$. $\diamondsuit$ - $N=10000$,
$L=50$. In Fig.~2b (for $L=20$) we show how $v_a$
changes if the noise is kept constant and the density is increased.

Fig. 3 Dependence of $\log v_a$ on $\log [(\eta_c(L)-\eta)/ \eta_c(L)] $
and $\log [(\rho-\rho_c(L)/ \rho_c(L)]$.  The slope of the lines fitted
to the data can be associated with the critical exponents $\beta$ and
$\delta$.  a) is for $\rho=0.4$, b) is for  $L=20$
and $\eta=2.0$

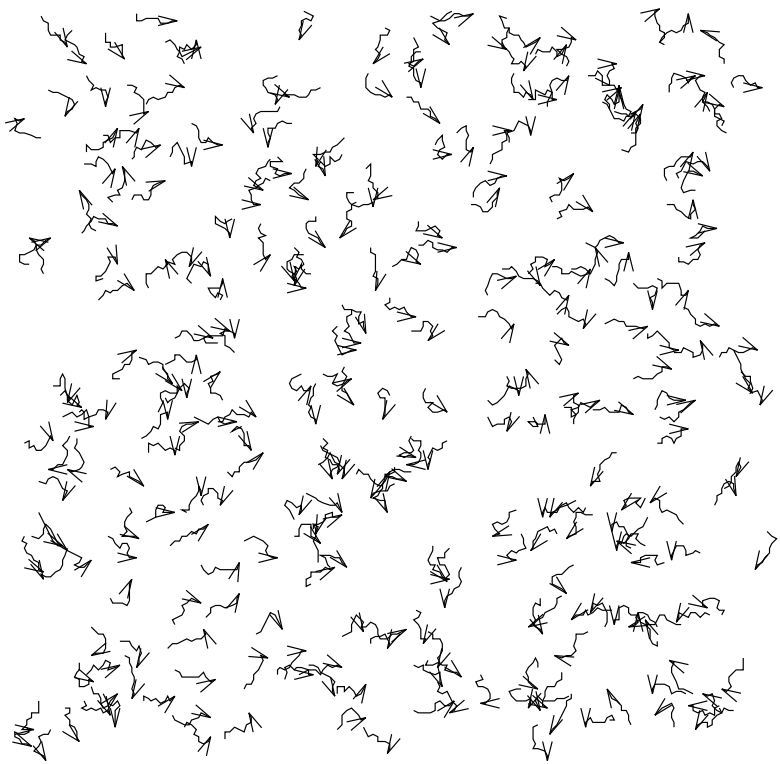
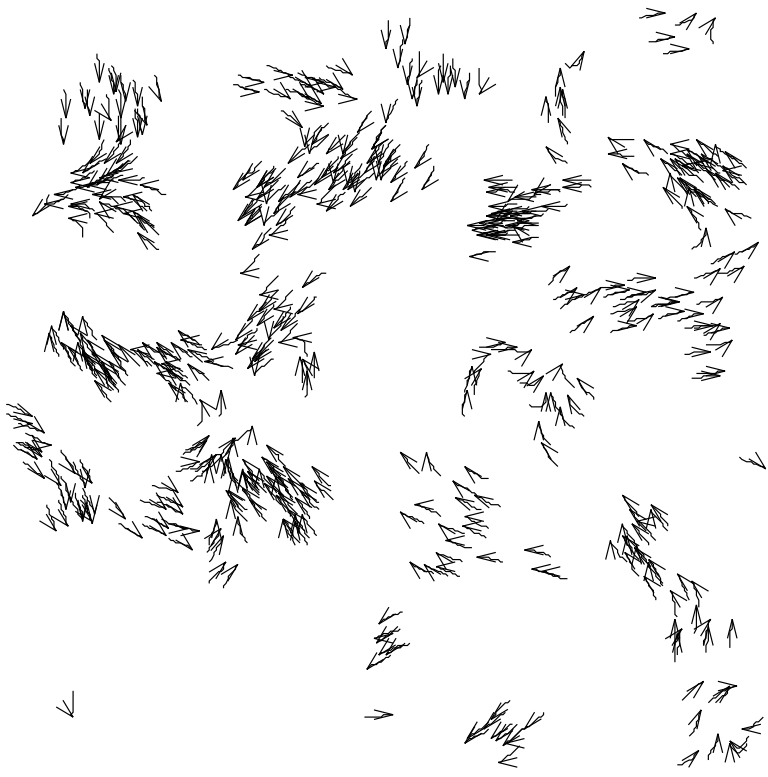
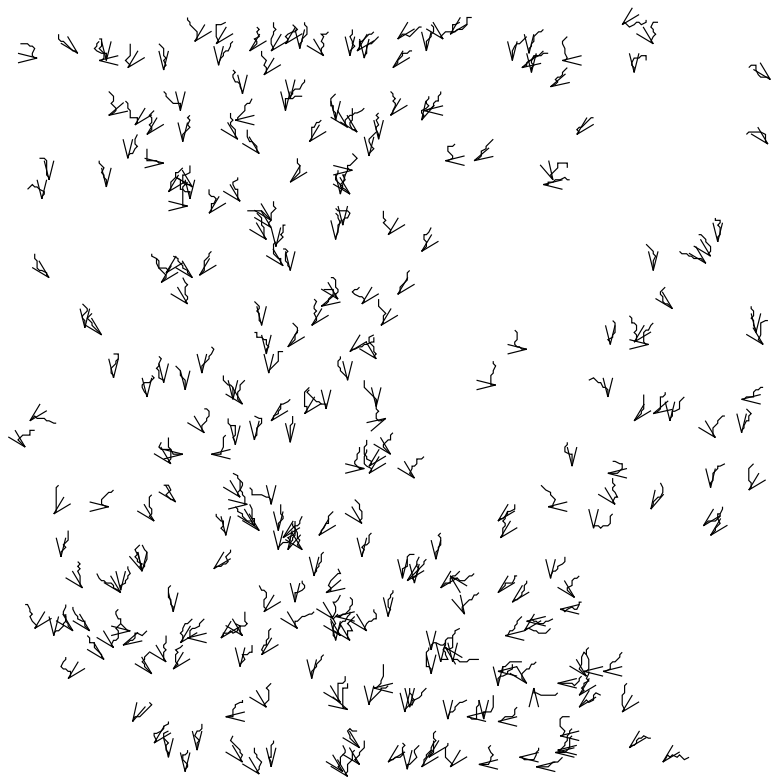
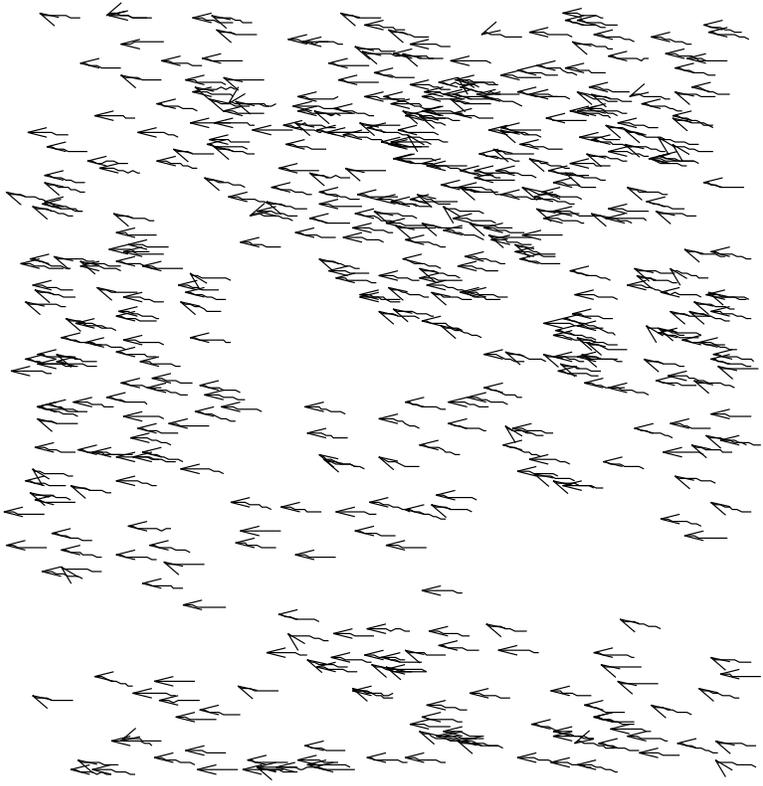
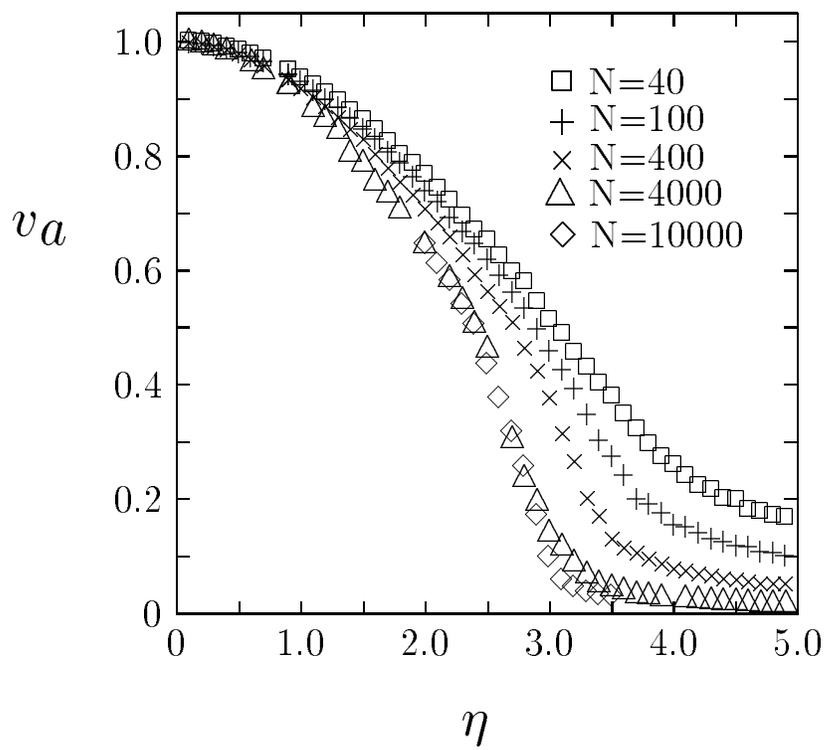
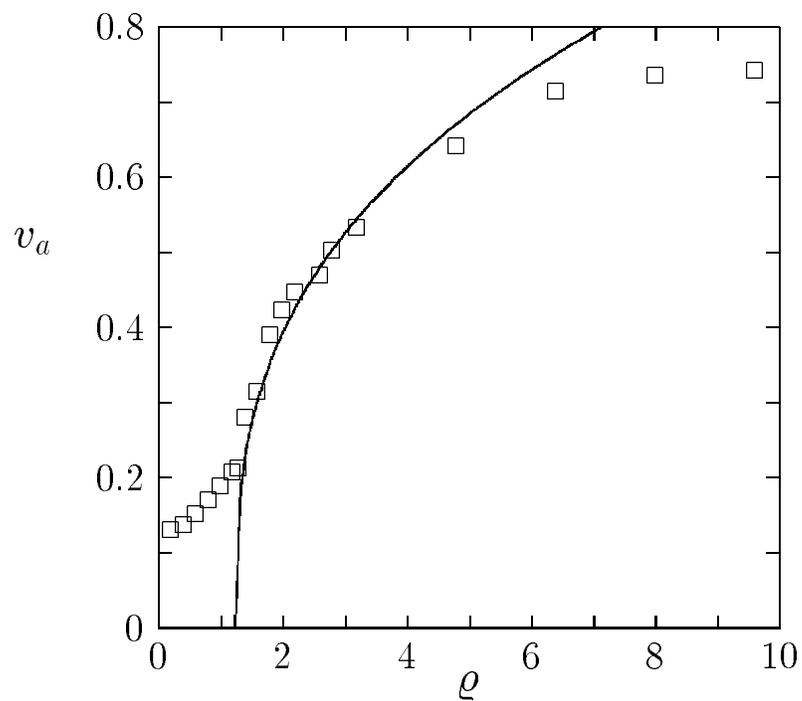
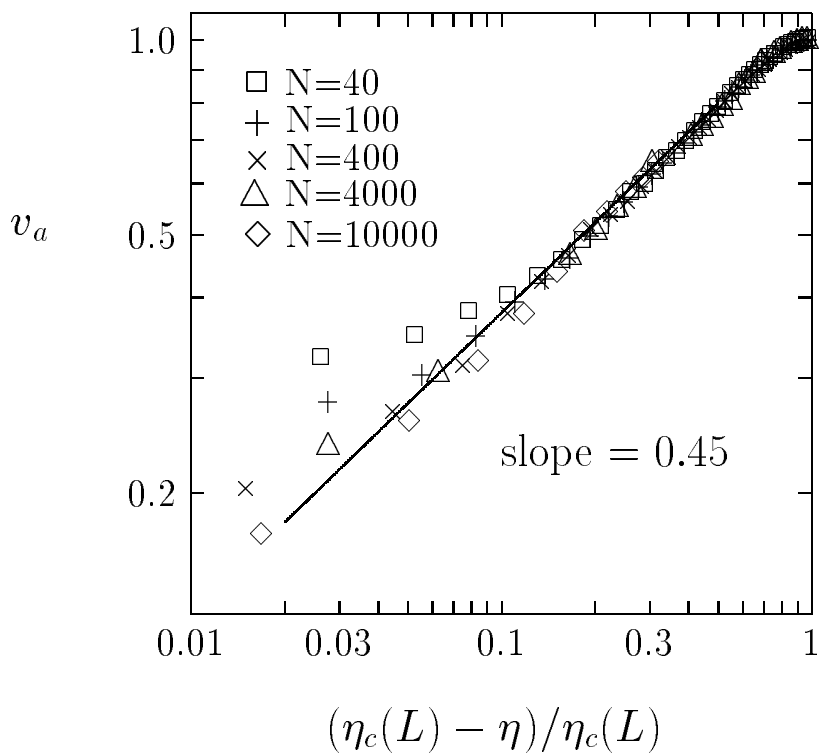
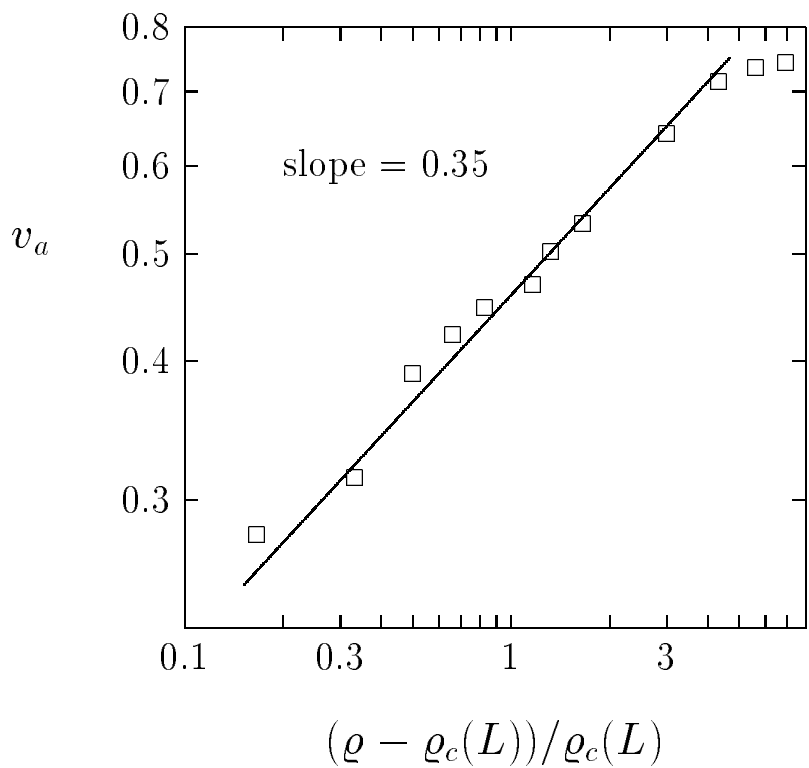
\bye